\documentclass[aps,prb,twocolumn, showpacs]{revtex4}

\usepackage{graphicx}
\usepackage{dcolumn}
\usepackage{bm}

\begin{document}

\title{Numerical study of spin quantum Hall transitions in superconductors with
broken time-reversal symmetry}

\author{Qinghong Cui$^1$, Xin Wan$^2$, and Kun Yang$^1$}

\affiliation{$^1$National High Magnetic Field Laboratory and Department of
Physics, Florida State University, Tallahassee, Florida 32306, USA}
\affiliation{$^2$Institut f\"ur Nanotechnologie, Forschungszentrum Karlsruhe,
76021 Karlsruhe, Germany}

\date{\today}

\begin{abstract}

We present results of numerical studies of spin quantum Hall transitions in
disordered superconductors, in which the pairing order parameter breaks
time-reversal symmetry. We focus mainly on $p$-wave superconductors in which
one of the spin components is conserved. The transport properties of the system
are studied by numerically diagonalizing pairing Hamiltonians on a lattice, and
by calculating the Chern and Thouless numbers of the quasiparticle states. We
find that in the presence of disorder, (spin-)current carrying states exist
only at discrete critical energies in the thermodynamic limit, and the
spin-quantum Hall transition driven by an external Zeeman field has the same
critical behavior as the usual integer quantum Hall transition of
non-interacting electrons. These critical energies merge and disappear as
disorder strength increases, in a manner similar to those in lattice models for
integer quantum Hall transition.

\end{abstract}

\pacs{74.40.+k, 73.43.Nq, 72.15.Rn}
\maketitle

%%%%%%%%%%%%%%%%%%%%%%%%%%%%%%%
%                             %
%  Section 1. Introduction    %
%                             %
%%%%%%%%%%%%%%%%%%%%%%%%%%%%%%%

\section{Introduction}

Transport properties of quasiparticles in unconventional superconductors have
been of strong interest to condensed matter physicists since the discovery of
high $T_c$ cuprate superconductors. The cuprates, which are $d$-wave
superconductors, support gapless nodal quasiparticle excitations; these nodal
quasiparticles dominate heat and spin transport at low temperatures. Some heavy
fermion superconductors are also known to have $d$-wave pairing. Another class
of unconventional superconductors that have been receiving increasing attention
are $p$-wave superconductors. While the $p$-wave pairing was originally found
in superfluid $^3$He,~\cite{vollhardt} interest in it has been renewed more
recently due to advances in two distinct systems. Firstly, fractional quantum
Hall effects at fillings factor $\nu = 5/2$~\cite{willett87} are believed to be
the manifestation of the $p$-wave pairing of composite fermions.~\cite{moore91,
greiter91, morf98, rezayi00, scarola00} Secondly, a growing number of results
on the unconventional superconductivity of Sr$_2$RuO$_4$ emerge in favor of a
triplet-pairing order parameter.~\cite{mackenzie03}

In addition to the experimental relevance, unconventional, disordered
superconductors are also of great interest for theoretical reasons, as they
represent new symmetry classes in disordered non-interacting fermion problems
that are not realized in metals. A classification of these symmetry classes
have been advanced recently.~\cite{altland97, zirnbauer96} Depending on the
existence (or the lack) of time-reversal and spin-rotation symmetries, dirty
superconductors can be classified into four symmetry classes, CI, DIII, C, and
D in Cartan's classification scheme. These classes are believed to complete the
possible universality classes~\cite{mehta} in disordered single-particle
systems.~\cite{altland97, zirnbauer96}

Systems of class C with broken time-reversal invariance but preserved
spin-rotation symmetry exhibit universal critical behavior which has been under
extensive investigation recently.~\cite{kagalovsky, gruzberg99, senthil99,
evers03, mirlin03} Such systems, which can be realized in two-dimensional
superconductors with $d_{x^2-y^2}+id_{xy}$ symmetry, have a distinct signature
of a critical density of states at criticality.~\cite{gruzberg99, senthil99}
Analogous to the conventional quantum Hall effect where the Hall conductance is
quantized, the Hall conductance of the spin current of such systems is
quantized. This effect is, therefore, called spin quantum Hall effect, because
spin, rather than charge, is conserved in such a superconductor.

Studies of the critical behavior of class D with both time-reversal and
spin-rotation symmetries broken have a longer history, starting from the
two-dimensional random-bond Ising model (RBIM).~\cite{nishimori81, singh96,
cho97, honecker01, read01, gruzberg01, merz02, chalker} The RBIM can be mapped
onto the Cho-Fisher network model,~\cite{cho97} which resembles the original
Chalker-Coddington network model~\cite{chalker88} for the integer quantum Hall
plateau transition, but with a distinct symmetry. The class D models may have
an even richer phase diagram; in addition to the spin (or thermal, if the
spin-rotation symmetry is completely broken) quantum Hall phase mentioned above
and an insulating phase (which is always possible), they may also support a
metallic phase.~\cite{senthil00}

The possibility of spin (or thermal) quantum Hall states in unconventional
superconductors allows one to draw a close analogy between the quantum Hall
effect and superconductivity, as well as to transfer theoretical or numerical
methods developed in the study of one system to another. One of the
well-developed numerical methods in the study of the quantum Hall effect is the
calculation of Chern numbers of either single- or many-electron states, which
allows one to distinguish between current carrying and insulating states
unambiguously, even in a finite-size system. The Chern number method has been
very successful in the studies of quantum Hall transitions, for both
integral~\cite{arovas, huo92, yang96, bhatt02} and fractional
effects,~\cite{sheng03} as well as in bilayer systems,~\cite{sheng03b} and also
in other contexts.\cite{shengweng, yangbhatt, yangmacd} For example, by
finite-size scaling, the localization length exponent $\nu \approx 2.3$ has
been obtained,~\cite{huo92, yang96, bhatt02} consistent with other
estimates.~\cite{huckestein95} More recent application of this method to
fractional quantum Hall states (where one inevitably has to deal with
interacting electrons) allows one to determine the transport gap
numerically,~\cite{sheng03} which is not possible from other known numerical
methods.

In this paper, we report results of numerical studies on a lattice model of
disordered $p$-wave superconductors with $p_x+ip_y$ pairing, which conserves
the $z$-component of electron spin. As we will show later, this is an example
of the class D model, and in certain sense the simplest model that supports a
spin quantum Hall phase. We study the localization properties of the
quasiparticle states by calculating the Chern and Thouless numbers of the
individual states, in ways similar to the corresponding studies in the quantum
Hall context mentioned above. Physically, the Chern and Thouless numbers
correspond to the Hall and longitudinal {\em spin} conductivities in the
present context, respectively, which are also related to the Hall and
longitudinal thermal conductivities. We note that while it has been pointed out
earlier that quasiparticle bands or individual quasiparticle states can be
labeled by their topological Chern numbers,\cite{hat} the present work
represents the first attempt to calculate them numerically and use them to
study the localization properties of the quasiparticle states in the context of
unconventional superconductors.

Our main findings are summarized as the following. We find that the $p$-wave
model we study supports an insulating phase and a spin quantum Hall phase with
spin Hall conductance one in appropriate unit. For relatively weak disorder,
there exist two critical energies at which current-carrying states exist,
carrying a total Chern number (or spin Hall conductance in proper unit) $+1$
and $-1$, respectively; they are responsible for the spin quantum Hall phase.
Phase transitions between these two phases may be induced either by changing
the disorder strength, or by applying and sweeping a Zeeman field. The
field-driven transition is found to have the same critical behavior as the
integer quantum Hall transition of non-interacting electrons. As disorder
strength increases the two critical energies both move toward $E=0$, and
annihilate at certain critical disorder strength, resulting in an insulating
phase in which all quasiparticle states are localized. No metallic phase is
found in our model.

The remainder of the paper is organized as the following. In section II we
introduce the model Hamiltonian of our numerical study, and discuss its
symmetry. We also describe the application of Chern and Thouless number methods
to the present problem in some detail. We present our numerical results in
section III, including results of the finite-size scaling analysis of the
numerical data. Section IV is reserved for a summary and the discussion of our
results.

%%%%%%%%%%%%%%%%%%%%%%%%%%%%%%%%%%%%%%%%%%%%%%
%                                            %
%  Section 2. Model and Numerical Methods    %
%                                            %
%%%%%%%%%%%%%%%%%%%%%%%%%%%%%%%%%%%%%%%%%%%%%%

\section{\label{sec:model}Model and Numerical Methods}

We consider electrons moving on a two-dimensional square lattice with linear
size $L$, in the presence of pairing and random potentials, described by the
Hamiltonian
\begin{eqnarray}
\mathcal{H} & = & - t \sum_{<i,j>}(c_{i\uparrow}^{\dagger} c_{j\uparrow}^{}
   + c_{i\downarrow}^{\dagger} c_{j\downarrow}^{})\nonumber\\
   & & + \sum_{<i, j>} (
   \Delta_{ij} c_{i\uparrow}^{\dagger} c_{j\downarrow}^{\dagger}
   + \Delta_{ij}^{\ast} c_{j\downarrow} c_{i\uparrow} )\nonumber\\
   & & + \sum_i (u_i - \mu) (c_{i\uparrow}^{\dagger} c_{i\uparrow}^{}
   + c_{i\downarrow}^{\dagger} c_{i\downarrow}^{}),
   \label{eq:hmlt}
\end{eqnarray}
where $u_i$ is the random potential on site $i$ which are independent random
variables distributed uniformly from $-W/2$ to $W/2$, and $\mu$ is the chemical
potential of the electrons. $t$ is the nearest neighbor hopping integral and we
choose $t = 1$ as the unit of energy from now on. We define, for a $p$-wave
superconductor, the pairing potential $\Delta_{j, j \pm e_x} = \pm \Delta$ and
$\Delta_{j, j \pm e_y} = \pm i\Delta$, where $e_x$ and $e_y$ are unit vectors
along $x$- and $y$-axis, respectively. For a finite-size system, we introduce
the generalized periodic boundary condition $c_{j + \mathbf{L}_i \uparrow} =
e^{i \theta_i} c_{j \uparrow}$, $c_{j + \mathbf{L}_i \downarrow} = e^{-i
\theta_i} c_{j \downarrow}$ \,($i = x, y$). We note that while the total spin
of the system is not conserved for the $p$-wave pairing, the $z$-component of
the spin is conserved due to our choice that pairing only occurs between
electrons with opposite spin. This becomes especially clear if we rewrite the
Hamiltonian in terms of particle-hole transformed operators for the electrons
with down spins:
\begin{equation}
d_{i\uparrow} = c_{i\uparrow}, \qquad
d_{i\downarrow} = c_{i\downarrow}^{\dagger},
\end{equation}
so that
\begin{equation}
\mathcal{H} = (d_{\uparrow}^{\dagger} \, d_{\downarrow}^{\dagger})
   \left( \begin{array}{cc}
   h & \Delta \\
   \Delta^{\dagger} & - h^T
   \end{array} \right) \left(\begin{array}{c}
   d_{\uparrow} \\ d_{\downarrow} \end{array} \right)
   \equiv d^{\dagger}\, H \, d,
\label{eq:hmlt2}
\end{equation}
where $h$ is the Hamiltonian without pairing potentials for each spin
component, and $\Delta = (\Delta_{mn})$. Clearly the number of $d$ particles is
conserved, reflecting the conservation of the $z$-component of the total
electron spin. Thus the corresponding transport properties of the $z$-component 
spin are well-defined; in the following we simply use the word spin to refer to 
its $z$-component, and spin conductances refer to the ratios between the 
$z$-component of the spin current and the gradient of the $z$-component of the 
Zeeman field.

The $p$-wave pairing symmetry $\Delta_{mn} = -\Delta_{nm}$ leads to a special
symmetry of the Hamiltonian:
\begin{equation}
\sigma_x H \sigma_x = - H^T,  \qquad
\sigma_x = \left(\begin{array}{cc}
   0 & {\bf 1} \\
   {\bf 1} & 0
   \end{array} \right), \label{symmetry}
\end{equation}
which makes the current problem distinct from the usual problem of electrons
moving in a random potential. In fact, it is an example of the symmetry class D
in the classification of Altland and Zirnbauer.\cite{altland97}

We note that the model we study here, Eq.~(\ref{eq:hmlt}), is not of the most
general form of symmetry class D\cite{altland97} for spin-$1\!/2$ fermions,
which, by classification, has no spin-rotation symmetry along any direction.
Consider a generic Hamiltonian for quasiparticles in a superconductor:
\begin{equation}
{\cal H} = \sum_{\alpha\beta} \left (
h_{\alpha\beta} c^\dagger_\alpha c_\beta
+ {1 \over 2} \delta_{\alpha\beta} c^\dagger_\alpha c^\dagger_\beta
+ {1 \over 2} \delta^*_{\alpha\beta} c_\beta c_\alpha
\right ),
\label{classD}
\end{equation}
where $\alpha$ and $\beta$ are indices that label both lattice site and spin of
the electron, running from $1$ to $2N$ (if $N$ is the number of lattice sites).
The Hamiltonian can be solved by the Bogoliubov transformation, or, explicitly,
by the diagonalization of the $4N \times 4N$ matrix
\begin{equation}
\hat{H} = \left (
\begin{array}{cc}
h & \delta \\
-\delta^* & -h^T
\end{array}
\right ).
\end{equation}
In the generic case of class D (without time-reversal and spin-rotation
symmetries), the only constraint on $\hat{H}$ is
\begin{equation}
\label{phsymmetry}
\hat{H}^\dagger = \hat{H} = -\Sigma_x \hat{H}^T \Sigma_x,
\end{equation}
where
\begin{equation}
\Sigma_x = \left (
\begin{array}{cc}
0 & {\bf 1}_{2N} \\
{\bf 1}_{2N} & 0
\end{array}
\right ) = \sigma_x \otimes {\bf 1}_{2N}.
\end{equation}
This constraint, which comes from both hermiticity and Fermi statistics, is the
same as Eq.~(\ref{symmetry}). Note that this Hamiltonian is twice as large as
the $p$-wave pairing Hamiltonian we study [Eq.~(\ref{eq:hmlt2})], although they
belong to the same symmetry class for the following reasons. Interestingly, the
partial spin-rotation symmetry along $z$-axis leads to a decomposition of the
$4N \times 4N$ matrix $\hat{H}$ into two homomorphic subblocks. One subblock
corresponds to spin-up particles and spin-down holes, and the other to spin-up
holes and spin-down particles. Without additional symmetry, each subblock
belongs to the conventional unitary ensemble.~\cite{lyanda-geller94} On the
other hand, if spin-rotation symmetry in other directions is present (as in
$d$-wave or other singlet pairing), the spin-up particles and spin-down holes
are equivalent, and the coupling between them is symmetric; this is, in fact,
the case of class C.  In the present case, however, the coupling between
spin-up particles and spin-down holes is {\em antisymmetric}, required by the
special $p$-wave pairing we introduce. Therefore, the model we study is
equivalent to pairing between spinless or spin-polarized fermions, also
described by a Hamiltonian of the form [Eq.~(\ref{classD})], with $\alpha$ and
$\beta$ label lattice sites only. It is in this sense that we can study
well-defined spin transport in a class D model.

It is also useful for us to consider the presence of a {\em uniform}
Zeeman field:
\begin{eqnarray}
H_B &=& \mu_B B \sum_m (c_{m\uparrow}^{\dagger} c_{m\uparrow}^{}
   - c_{m\downarrow}^{\dagger} c_{m\downarrow}^{})
\nonumber\\
&=& \mu_B B \sum_m (d_{m\uparrow}^{\dagger} d_{m\uparrow}^{}
   + d_{m\downarrow}^{\dagger} d_{m\downarrow}^{})+const.
\end{eqnarray}
We note that the Zeeman field plays a role of the Fermi energy for the
(conserved) $d$ particles. More importantly, its presence changes the symmetry
property of the systems, because $H_B$ does {\em not} obey
Eq.~(\ref{symmetry}).

The spin Hall conductance of an individual quasiparticle
eigenstate $|m\rangle$ can
be calculated by the Kubo formula
\begin{equation}
\sigma_{xy}^S (m) = \frac{i \hbar}{A} \sum_{n \ne m}
   \frac{\langle m|j_{x}^{S}|n\rangle \langle n|j_{y}^{S}|m \rangle
   - \langle m|j_{y}^{S}|n \rangle \langle n|j_{x}^{S}|m\rangle}
   {(E_n - E_m)^2},
\end{equation}
where $A = L^2$ is the area of the system. $|m\rangle$, $|n\rangle$ are
quasiparticle eigenstates of the Hamiltonian [Eq.~(\ref{eq:hmlt})] and
$j_{x}^{S}$, $j_{y}^{S}$ the components of the spin current operator. Following
Thouless and co-workers,~\cite{thouless1982,thouless1985} we can show that the
spin Hall conductance averaged over boundary conditions is related to a
topological quantum number:
\begin{eqnarray}
\langle \sigma_{xy}^S (m) \rangle
   &=& \frac{\hbar}{8 \pi} \int\!\!\!\int d\theta_x d\theta_y
   \frac{1}{2\pi i}\bigg [
   \left \langle \frac{\partial m } {\partial \theta_y}
   \bigg| \frac{\partial m}{\partial \theta_x} \right\rangle \nonumber \\
   & & \qquad - \left\langle \frac{\partial m}{\partial \theta_x}
   \bigg| \frac{\partial m}{\partial \theta_y}
   \right \rangle \bigg ]\nonumber \\
   &=& \frac{\hbar}{8 \pi} C_1(m) , \label{eq:chern}
\end{eqnarray}
where $C_1(m)$ is an integer and known as the first Chern index. As is widely
used in quantum Hall transitions and other contexts,\cite{arovas, huo92,
yang96, bhatt02, sheng03, sheng03b, shengweng, yangbhatt, yangmacd} $C_1(m)$
can be used to distinguish current carrying states from localized states
unambiguously, even in finite-size systems, thus providing a powerful method to
study the localization properties of the quasiparticle states.

%%%%%%%%%%%%%%%%%%%%%%%%%%%%%%%%%%
%  Introducing Thouless number
%%%%%%%%%%%%%%%%%%%%%%%%%%%%%%%%%%

An alternative way to study the localization properties of the states is to
calculate the Thouless number (also known as the Thouless conductance) of the
states at a given Fermi energy $E$, defined as~\cite{thouless1972,
thouless1975}
\begin{equation}
g_T(E) = \frac{\langle |\delta E| \rangle}{\Delta E}
   \sim \frac{8 \pi}{\hbar} \sigma_{xx}^S,
\end{equation}
where $\Delta E $ is the average energy level spacing at energy $E$, and
$\langle |\delta E| \rangle$ is the average energy level shift caused by the
change of the boundary condition from periodic to anti-periodic in one spatial
direction. It was argued in the context of electron localization that $g_T(E)$
is proportional to the longitudinal conductance of the system;
\cite{thouless1972, thouless1975} in the present context we expect it to
provide a measure of the longitudinal spin conductance of the superconductor.
Thouless numbers have also been numerically studied for the conventional
integer quantum Hall effect, in both full~\cite{yang00} and
projected~\cite{wan01} lattice models.

In this work we carry out numerical calculations to diagonalize the Hamiltonian
$H$ to obtain the exact quasiparticle eigen wave functions. We calculate their
Chern and Thouless numbers to study their localization properties, and perform
finite-size scaling analysis to extract critical behavior of the transitions
driven by the change of the disorder strength $W$ or the Zeeman field.

%%%%%%%%%%%%%%%%%%%%%%%%%%%%%%%%%%
%                                %
%  Section 3. Numerical Results  %
%                                %
%%%%%%%%%%%%%%%%%%%%%%%%%%%%%%%%%%

\section{\label{sec:results}Numerical Results}

In the absence of the random potential we can diagonalize the Hamiltonian
[Eq.~(\ref{eq:hmlt})] in the momentum space, and the energy spectrum is
\begin{equation}
E_k = \sqrt{\varepsilon_{k}^{2} + |\Delta_{k}|^{2}}, \label{spectrum}
\end{equation}
where $\varepsilon_k = -2 t (\cos k_x + \cos k_y)-\mu$ is the single-particle
kinetic energy, and $\Delta_k = 2i\Delta(\sin k_x + i\sin k_y)$ the $p$-wave
pairing order parameter. From Eq.~(\ref{spectrum}), we expect that there is an
energy gap between two bands, which should be stable against weak disorder,
while the gap will be closed when the disorder becomes strong enough.

\begin{figure}
\includegraphics[width = 0.5\textwidth]{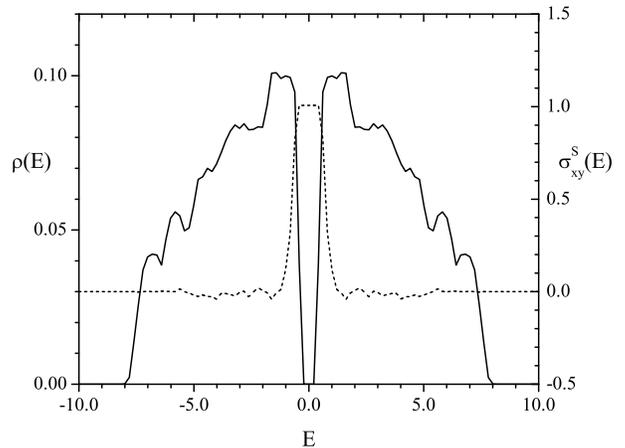}
\caption{\label{fig:pcnh} Density of states $\rho(E)$ (solid line) and spin
Hall conductance $\sigma^S_{xy}(E)$ (dotted line, in units of $\hbar / 8 \pi$)
for $L = 10$ and $W = 4.0$. We average over 500 samples of different random
potential realizations.}
\end{figure}

In this work, we choose $\Delta = 0.5$ and the chemical potential $\mu = 3.0$
to avoid the van Hove singularity at zero energy in the single electron
spectrum. To calculate the Chern number of each eigenstate, we evaluate the
integral in Eq.~(\ref{eq:chern}) numerically over the boundary phase space $0
\le \theta_x ,\, \theta_y \le 2\pi$. We divide the boundary phase space into $M
\times M$ square grids with $M = 20$-80, depending on the system size $L =
10$-40 to achieve desired precision. Figure~\ref{fig:pcnh} shows the density of
states (DOS) (per lattice site and spin species) $\rho(E)$ for a system with $L
= 10$ and $W = 4.0$. For such a relatively weak disorder, the superconducting
gap is still visible. Also shown is the spin Hall conductance $\sigma^S_{xy}$
as a function of quasiparticle Fermi energy $E=\mu_BB$, calculated by summing
up Chern number of states below the Fermi energy. We find that $\sigma^S_{xy}$
jumps from zero up by one unit near the (disorder-broadened) lower band edge,
and jumps back to zero above the gap. Therefore, a plateau in $\sigma^S_{xy}$
is well developed around $E = 0$, clearly indicating the existence of a spin
quantum Hall phase. This phase with topological Chern number equal to one is
the simplest possible spin quantum Hall phase for non-interacting
quasiparticles; it is simpler, for example, than the corresponding phase of an
$d_{x^2-y^2}+id_{xy}$ superconductor, which carries a total Chern number two.

\begin{figure}
\includegraphics[width = 0.5\textwidth]{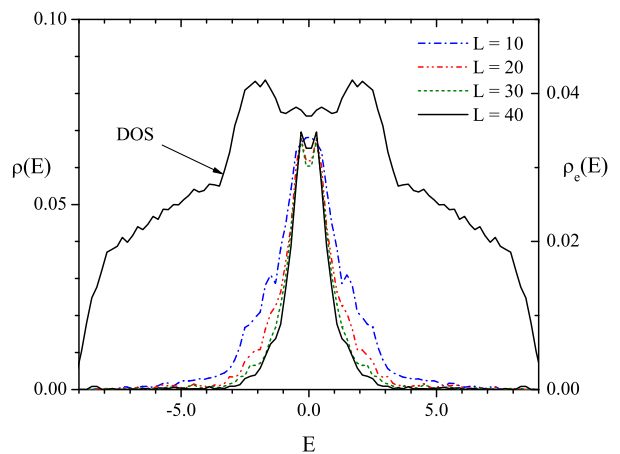}
\caption{\label{fig:pcnd} Density of states (DOS) $\rho(E)$ and density of
current carrying states (with nonzero Chern number), $\rho_e(E)$, for systems
with $L=10$-40 and $W = 8.0$.}
\end{figure}

\begin{figure}
\includegraphics[width = 0.5\textwidth]{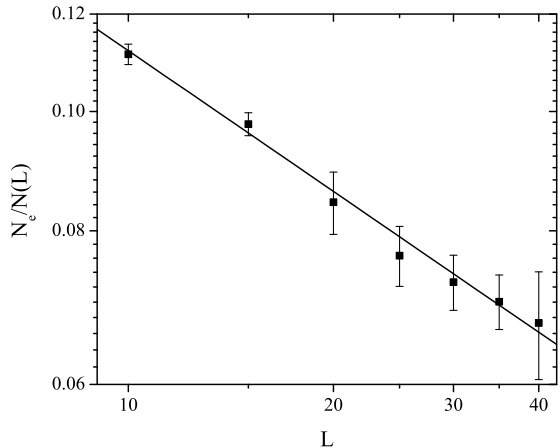}
\caption{\label{fig:pcns} Percentage of current carrying states $N_e/N(L)$
versus system size $L$ on a log-log scale for $W=8.0$. The solid line is a
power-law fit of the data.}
\end{figure}

In the following discussion, we focus on cases with disorder strong enough to
close the gap, and look for transitions from the spin quantum Hall phase to
other possible phases, driven by either the disorder strength $W$ or the
quasiparticle Fermi energy. In Fig.~\ref{fig:pcnd}, we plot the total DOS
$\rho(E)$ (which is roughly system size independent) and the density of current
carrying states (defined as states with non-zero Chern number) $\rho_e(E)$ for
systems with $L = 10$-40. We find that $\rho_e(E)$ has a weak double-peak
structure near $E = 0$ for large $L$, whose width shrinks as $L$ increases.
This behavior is reminiscent of those seen in the numerical study of current
carrying states in the integer quantum Hall effect,\cite{huo92, yang96} where
the current carrying states exist only at discrete critical energies in the
thermodynamic limit and, thus, the width of $\rho_e(E)$ peak(s) shrinks to zero
as $L$ increases toward infinity. In the present case the two peaks correspond
to two such critical energies, carrying a total Chern number $+1$ and $-1$,
respectively, which are responsible for the spin quantum Hall plateau when the
Fermi energy is between them (so that only the lower critical energy is below
the Fermi energy). According to the scaling theory of localization, $\rho_e(E)$
depends on $L$ only through a dimensionless ratio $L/\xi(E)$ when the system
size becomes sufficiently large; the localization length $\xi$ diverges in the
vicinity of a critical energy $E_c$ as $\xi(E) \sim |E - E_c|^{- \nu}$.
Therefore, the number of current carrying states $N_e(L)$ behaves as
\begin{equation}
\label{cnscaling}
N_e(L) = 2L^2 \int_{-\infty}^{\infty} \rho_e(E) dE \sim L^{2 - 1/\nu},
\end{equation}
from which we can estimate $\nu$. Assuming we have a similar situation here, we
plot $N_e(L)$, normalized by the total number of states $N(L)=2L^2$, on a
log-log scale in Fig.~\ref{fig:pcns}. Just as in the quantum Hall
case,\cite{huo92,yang96}, we can fit the data to a power law (a straight line
in the log-log plot) as in Eq.~(\ref{cnscaling}) reasonably well, and obtain
\begin{displaymath}
\nu = 2.6 \pm 0.2.
\end{displaymath}
This is close to the corresponding exponent $\nu = 2.3 \pm 0.1$ for the integer
quantum Hall transition. These results suggest that just as in the case of the
integer quantum Hall effect, current carrying states exist at discrete critical
energies in the thermodynamic limit, and the spin quantum Hall transition
driven by the Zeeman field (or equivalently, the quasiparticle Fermi energy)
has the same critical behavior as the integer quantum Hall transition. This is
expected on the symmetry ground, because in this case the critical energies are
away from $E=0$, and thus can only be reached in the presence of the Zeeman
field. As discussed earlier, the Zeeman field breaks the symmetry of
Eq.~(\ref{symmetry}) and reduces the symmetry of the present problem to that of
electrons moving in a magnetic field and a random potential.

\begin{figure}
\includegraphics[width = 0.5\textwidth]{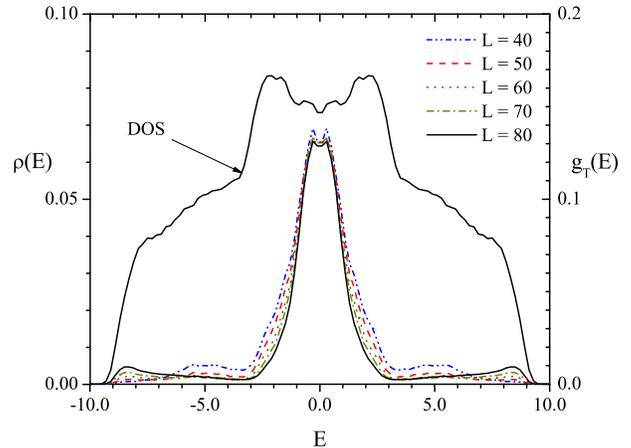}
\caption{\label{fig:ptnd8} Density of states $\rho(E)$ and Thouless number
$g_T(E)$ for systems with $L=40$-80 and $W = 8.0$.}
\end{figure}

\begin{figure}
\includegraphics[width = 0.5\textwidth]{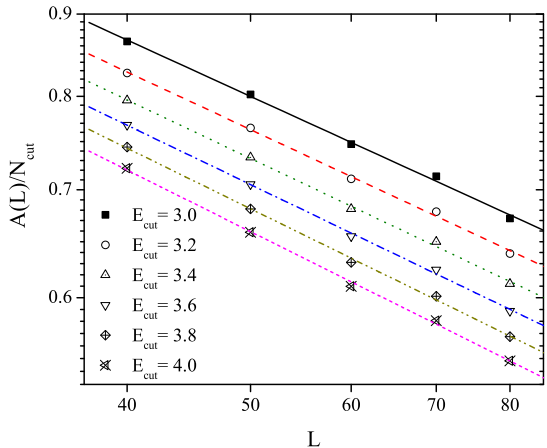}
\caption{\label{fig:ptns8} Area $A(L)$ of Thouless number $g_T(E)$ normalized
by number of states $N_{\text{cut}}$ counted, versus system size $L$ on a
log-log scale for different cutoff energy $E_{\text{cut}}$ and $W = 8.0$. The
lines are power-law fits of the data.}
\end{figure}

While the Chern numbers measure the ability of individual states to carry spin
Hall current, we have also calculated the Thouless conductance $g_T (E)$, which
is a measure of the longitudinal spin conductance. Unlike the Chern number
calculation which requires the diagonalization of the Hamiltonian for many
different boundary conditions, the Thouless number calculation only needs the
diagonalization at two different boundary conditions, thus allowing us to study
larger systems. On the other hand, it is known in the numerical study of
quantum Hall effect that Chern number calculation reaches the scaling behavior
at smaller system sizes. Therefore, these two methods are complementary to each
other. Figure~\ref{fig:ptnd8} shows $\rho(E)$ and $g_T (E)$ for systems with $L
= 40$-80, and with $W = 8.0$.  We find that $g_T (E)$ has a similar double-peak
structure as $\rho_e(E)$ with peaks locating at the same energies, and that the
peaks become narrower as $L$ increases. In the following we perform the similar
scaling analysis based on the zeroth moment of $g_T (E)$ as for the Chern
numbers. Namely, we compute the area $A(L)$ under $g_T(E)$ and expect
\begin{equation}
A(L) = \int_{-\infty}^{\infty} g_T(E) dE \sim L^{- 1/\nu}.
\end{equation}
One slight complication is that unlike $\rho_e(E)$, $g_T (E)$ has long tails
extending to the edges of $\rho(E)$, which clearly has no connection to the
critical behavior near the critical energies. To eliminate the influence of
these artificial tails, we introduce a cutoff energy $E_{\text{cut}}$, and
exclude contributions from $|E| > E_{\text{cut}}$. Based on the Chern number
calculation above (Fig.~\ref{fig:pcnd}), as well as the $g_T(E)$ curves
themselves, we can safely choose $E_{\text{cut}}$ between $3.0$ and $4.0$,
beyond which we find essentially no current carrying states for $L \ge 40$. In
Figure~\ref{fig:ptns8}, we plot, on a log-log scale, the area $A(L)$ normalized
by the area under the DOS curve between $-E_{\text{cut}}$ and $E_{\text{cut}}$:
\begin{displaymath}
N_{\text{cut}} = \int_{-E_{\text{cut}}}^{E_{\text{cut}}} \rho(E; L) dE
\end{displaymath}
for a series of different $E_{\text{cut}}$, and list the corresponding $\nu$ in
Table~\ref{tab:ptns8}. We find that $\nu$ has very weak dependence on the
choice of the cutoff energy and its variation between $2.54$ and $2.79$ is
consistent with the results obtained from the Chern number calculation.

\begin{table}
\begin{ruledtabular}
\begin{tabular}{clllllllllll}
$E_{\text{cut}}$ &
3.0 & 3.2 & 3.4 & 3.6 & 3.8 & 4.0 \\
$\nu$ & 2.79 & 2.73 & 2.68 & 2.64 & 2.58 & 2.54 \\
$\delta\nu$ & 0.08 & 0.08 & 0.08 & 0.07 & 0.07 & 0.07 \\
\end{tabular}
\end{ruledtabular}
\caption{\label{tab:ptns8} Critical exponent $\nu$ for different cutoff
energy $E_{\text{cut}}$ with $W = 8.0$.}
\end{table}

\begin{figure}
\includegraphics[width = 0.5\textwidth]{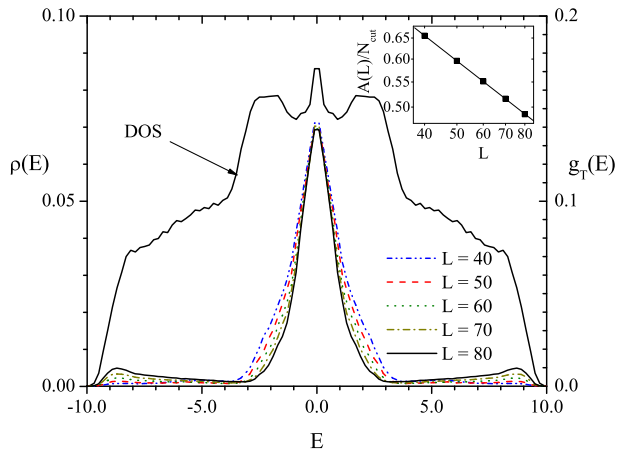}
\caption{\label{fig:ptnd9} Density of states $\rho(E)$ and Thouless number
$g_T(E)$ for systems with $L=40$-80 and $W = 9.0$. The inset shows the area of
Thouless number $g_T(E)$ divided by $N_{\text{cut}}$ for $E_{\text{cut}} =
4.0$.}
\end{figure}

We also studied other disorder strengths. In the case of the integer quantum
Hall transition,\cite{yang96} it is known that as the disorder strength
increases, the critical energies that carry opposite Chern numbers move close
together, merge, and disappear at some critical disorder strength $W_c$. In the
present case, we expect the same to happen and due to the symmetry of the
Hamiltonian, the critical energies can only merge at $E=0$. We present the
results for $W = 9.0$ in Fig.~\ref{fig:ptnd9}. In this case we no longer see
two split critical energies, suggesting that the two critical energies that
were clearly distinguishable at $W=8.0$ either (i) have moved too close to be
distinguishable at the accessible system sizes, or (ii) have just merged. We
believe scenario (i) is much more likely than (ii) based on the following
observations. (a) We find that the peak value of $g_T(E)$ is {\em independent}
of system size and takes the {\em same} value as that of $W=8.0$. (b) We have
performed the same scaling analysis of $g_T(E)$ as we did above for $W=8.0$ and
obtained a similar exponent $\nu\approx 2.3$ (see inset of
Fig.~\ref{fig:ptnd9}), which is even closer to the known value of the integer
quantum Hall transition. However, there is another possibility that instead of
entering the insulating phase (in which all quasiparticles states are
localized) immediately, the system is in a {\em metallic phase}, after the two
critical energies merge so that the system is no longer in the spin quantum
Hall phase. Senthil and Fisher\cite{senthil00} suggested that in this phase
both the DOS $\rho(E)$ and the conductance diverge logarithmically at the band
center. Interestingly, we indeed find $\rho(E)$ to be enhanced at $E=0$. We
believe, however, this is not associated to the metallic phase for the
following reasons. (i) No such enhancement is seen in the Thouless number,
which is a measure of the longitudinal conductance. (ii) We find $\rho(E)$ to
be essentially system size {\em independent} between $L=40$ and $L=80$, even at
$E=0$, while one expects\cite{senthil00} $\rho(L)\sim \log L$ in the metallic
phase. (iii) We find that (see below) the enhancement of $\rho(E)$  at $E=0$ is
also present at stronger disorder when the system is clearly insulating. Thus
it appears unlikely that the metallic phase is responsible for the single peak
in $g_T(E)$.

\begin{figure}
\includegraphics[width = 0.5\textwidth]{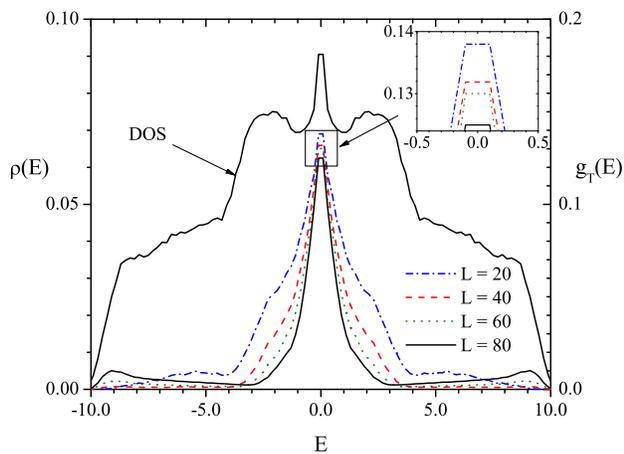}
\caption{\label{fig:ptnd10} Density of states $\rho(E)$ and Thouless number
$g_T(E)$ for systems with $L=40$-80 and $W = 10.0$. The inset is a blow-up of
the Thouless number curves near $E=0$, which shows that $g_T(E=0)$ decreases
with increasing $L$.}
\end{figure}

The situation is quite different as $W$ further increases. In
Fig.~\ref{fig:ptnd10}, we present results for $W=10.0$ and see a very different
behavior. Here the peak value of $g_T(E)$ systematically decreases as the
system size increases, exhibiting a characteristic insulating behavior.
Combined with results of smaller $W$, we conclude that in the absence of the
Zeeman field (or when the quasiparticle Fermi energy is at $E=0$), the system
is driven into the insulating phase from the spin quantum Hall phase as the
disorder strength $W$ increases. The critical strength $W_c$ is slightly above
$9.0$ and clearly below $10.0$. No evidence has been found for the existence of
an intermediate metallic phase that separates these two phases for our choice
of model parameters ($\mu=3.0$, $\Delta=0.5$, etc.).

The critical behavior of the transition driven by increasing $W$ is expected to
be different from the one driven by changing the Zeeman field discussed above,
due to the additional symmetry. In order to study the critical property one
first needs to determine the critical disorder strength $W_c$ accurately, which
we are unable to do within the accessible system size in our study. It would be
of significant interest to study this transition with more powerful computers
and/or other computational methods.

\begin{figure}
\includegraphics[width = 0.5\textwidth]{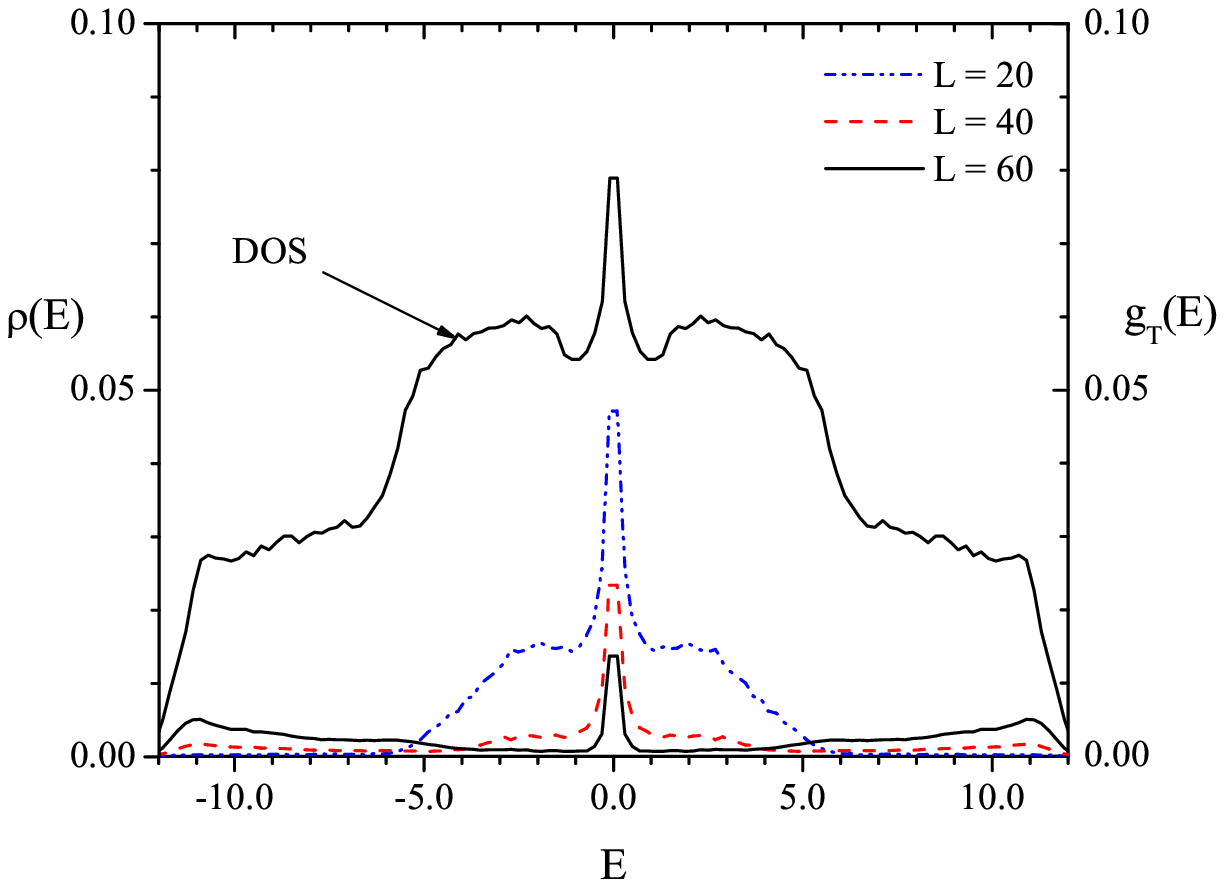}
\caption{\label{fig:ptnd15} Density of states $\rho(E)$ and Thouless number
$g_T(E)$ for systems with $L=20$-40 and $W = 15.0$.}
\end{figure}

\begin{figure}
\includegraphics[width = 0.5\textwidth]{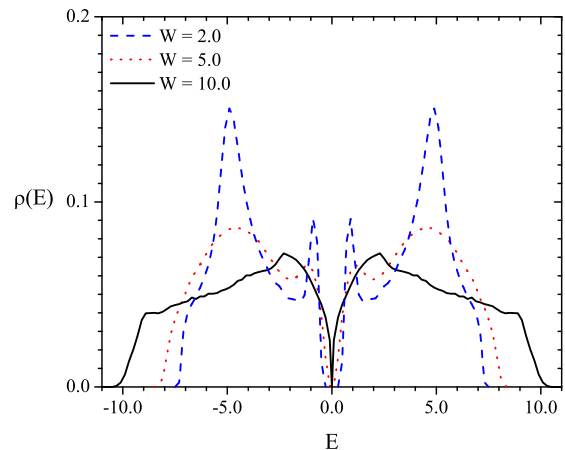}
\caption{\label{fig:ddd60w} Density of states of $d_{x^2-y^2}+id_{xy}$
superconductor with $L=60$, $\Delta_{x^2-y^2} = 1$, $\Delta_{xy} = 0.6$. We
average over 80 samples of different random potential realizations.}
\end{figure}

We give the results of $W = 15.0$ in Fig.~\ref{fig:ptnd15} as an example of
strong disorder, where all states are clearly localized. Here, the Thouless
number drops rapidly as the system size increases as expected. Interestingly,
the enhancement of the DOS at $E=0$ remains to be quite pronounced, suggesting
that it is {\em not} associated with possible metallic behavior discussed
above. For comparison, we have also calculated the DOS for a
$d_{x^2-y^2}+id_{xy}$ superconductor, by choosing the pairing order parameter
to be $\Delta_{j, j+e_x} = -\Delta_{j, j+e_y} = \Delta_{x^2-y^2}$, $\Delta_{j,
j+e_x+e_y} = -\Delta_{j, j +e_x-e_y} = i\Delta_{xy}$. In the $d$-wave
superconductor the total spin of the system is conserved, due to the singlet
nature of the pairing. As a consequence it belongs to the symmetry class C in
the classification by Altland and Zirnbauer.\cite{altland97} This model has
been studied in considerable detail in Refs. \onlinecite{kagalovsky,
gruzberg99, senthil99, evers03}. We plot the DOS for a $d_{x^2-y^2}+id_{xy}$
superconductor for different values to $W$ in Fig.~\ref{fig:ddd60w}. While the
gap vanishes just like the $p$-wave case for sufficiently large $W$, the DOS
exhibits a pseudogap behavior at $E=0$ for large $W$, in the vicinity of which
the DOS vanishes in an (apparently sublinear) power law as
predicted.\cite{gruzberg99, senthil99} This is a good example that the change
of symmetry profoundly affects the critical behavior as well as other
properties of the system.

\section{Discussion and Summary}

In this paper we have studied the localization properties of the quasiparticle
states in superconductors with spontaneously broken time-reversal symmetry,
which support spin quantum Hall phases. Our study is based on the exact
diagonalization of microscopic lattice models and the consequent numerical
calculation of the Chern and Thouless numbers of the quasiparticle states. Our
microscopic study is complementary to previous numerical work on this subject,
which have been based almost exclusively on effective network models with
appropriate symmetries.\cite{kagalovsky, senthil99, evers03, mirlin03, chalker,
cho97}

We have focused mostly on a $p$-wave pairing model in which the time-reversal
symmetry is broken by the (complex) pairing order parameter, while the
$z$-component of the total spin is conserved so that the transport properties
of the $z$-component of the spin is well defined. We find the system supports a
spin quantum Hall phase with spin Hall conductance one in appropriate unit, and
an insulating phase. Transitions between these two phases may be induced either
by changing the disorder strength, or by applying and sweeping a Zeeman field.
The field-driven transition is found to have the same critical behavior as the
integer quantum Hall transition of non-interacting electrons as expected on
symmetry grounds. The disorder-driven transition in the absence of the Zeeman
field is expected to have different critical properties due to additional
symmetry of the Hamiltonian. However, we have not been able to study the
critical behavior of this transition.

The symmetry properties of the $p$-wave pairing model in the absence of the
Zeeman field belongs to class D in the classification of general fermion
pairing models of Altland and Zirnbauer.\cite{altland97} It has been suggested
that in addition to the quantum Hall and the insulating phases, class D models
may also support a metallic phase,\cite{senthil00} which has logarithmically
divergent density of states and conductance. Such a system can have either a
direct transition between the quantum Hall and the insulating phases, or a
metallic phase separating these two phases. In our model we find a direct
transition between the spin quantum Hall and insulating phases, but no
definitive evidence for a metallic phase. This is not unusual as it is
known\cite{chalker} that specific microscopic models may or may not support the
metallic phase.

For comparison, we have also calculated the density of states of a $d$-wave
superconductor with $d_{x^2-y^2}+id_{xy}$ pairing order parameter, which
supports a spin quantum Hall phase with spin Hall conductance two in the same
unit. This model has different symmetry properties and belongs to class C in
the classification of Altland and Zirnbauer. We find that the density of states
vanishes with sublinear power law near $E=0$, in agreement with earlier
studies.\cite{gruzberg99, senthil99, evers03, mirlin03} This is in sharp
contrast to the $p$-wave case in which we observe an {\em enhanced} density of
states at $E=0$ for sufficiently strong disorder, demonstrating the profound
effect of symmetries on the low-energy properties of the system. While this
enhancement is somewhat reminiscent of the divergent density of states of the
possible metallic phase, further analysis suggests this is not the case. The
origin of this enhancement is currently unclear.

Finally we note that recently there is interest in the spin Hall effect in 
semiconductors with spin-orbit coupling, which is driven
by an electric field.~\cite{murakami,sinova} Physically this effect is quite 
different from the spin Hall effect discussed here, in the following ways.
(i) Our spin Hall effect is induced by the gradient of a Zeeman field that 
couples to spin, while the other effect is 
induced by an electric field that couples to charge. 
(ii) The existence of the spin Hall effect in our case
relies on the broken time-reversal symmetry in the pairing Hamiltonian, while
the time-reversal symmetry is intact in the Hamiltonians used in 
Refs.~\onlinecite{murakami,sinova}; instead the spin-Hall effect is present
due to the presence of spin-orbit coupling. In Ref.~\onlinecite{sinova}, a 
universal spin Hall conductance was found in a clean system; it is not clear
at present if this value has a topological origin as in our case, and
how stable this result is in the presence of disorder.

\acknowledgments We thank F. Evers, A. Mildenberger, and R. Narayanan for
helpful discussions, and A.~D. Mirlin for bringing
Ref.~\onlinecite{lyanda-geller94} to our attention. This work was supported by
NSF grant No. DMR-0225698 (Q.C. and K.Y.), the Research Corporation (Q.C.), and
the Schwerpunktprogramm ``Quanten-Hall-Systeme'' der Deutschen
Forschungsgemeinschaft (X.W.).

\end{document}